\documentclass[prd,aps,twocolumn]{revtex4}
\usepackage{amssymb}
\usepackage{graphicx}
\def\gr{$\gamma$-ray}
\begin{document}

\title{Neutrinos from Extra-Large Hadron Collider in the Milky Way} 
\author
{Andrii Neronov$^{1\ast}$, Dmitry Semikoz$^{2\dagger}$}

\affiliation{$^{1}$ ISDC, Astronomy Department, University of Geneva,
Ch. d'Ecogia 16, Versoix,1290, Switzerland\\
$^{2}$ Astroparticules et Cosmologie, 10 rue Alice Domon et Leonie Duquet, 
F-75205 Paris Cedex 13, France\\
E-mail:  $^\ast$Andrii.Neronov@unige.ch; $^\dagger$dmitri.semikoz@apc.univ-paris7.fr
}
\begin{abstract}
Neutrino telescope IceCube has recently discovered astrophysical neutrinos with energies in the TeV -- PeV range.  We use the data of Fermi \gr\ telescope to demonstrate that the neutrino signal has significant contribution from the Milky Way galaxy. Matching the \gr\ and neutrino spectra  we find that  TeV-PeV Galactic cosmic rays form a powerlaw spectrum with the slope $p\simeq 2.45$. This spectral slope is consistent with the average cosmic ray spectrum in the disks of the Milky Way and Large Magellanic Cloud galaxies. It is also consistent with the theoretical model of cosmic ray injection by diffusive shock acceleration followed by escape through the Galactic magnetic field with Kolmogorov turbulence. The locally observed TeV-PeV cosmic ray proton spectrum is softer than the average Galactic cosmic ray spectrum. This could be readily explained by variability of injection of cosmic rays in the local interstellar medium over the past $10^7$~yr and discreetness of the cosmic ray source distribution. 
\end{abstract}

\maketitle
\section{Introduction}

Our knowledge of the properties of cosmic rays produced by the star formation in the Milky Way galaxy is both precise and largely incomplete. It is precise because the details of the cosmic ray spectrum and elemental composition are measured with high precision \citep{pdg}. And it is incomplete because all the precision measurements refer to only single location within the Galaxy: the position of the Solar system. It is even not obvious what kind of information is contained in such single point measurements.  Deflections of cosmic rays by turbulent component of Galactic magnetic field forces them into a random walk through the interstellar medium (ISM) \citep{Berezinsky_book}. As a result, cosmic rays arrive from random directions on the sky. It is clear that some information on the sources is encoded in the  cosmic ray spectrum, mass composition and global anisotropy \citep{pdg}. However, it is not clear how the spectral  slope $p$ and the break energies of the locally measured   piecewise powerlaw cosmic ray spectrum $dN/dE\propto E^{-p}$ are related to the properties of source population(s) and their distribution in the Galaxy. The diffusion through the ISM and escape from the Galactic Disk modify the cosmic ray spectrum, in a model-dependent way.  The locally measured properties of the spectrum are not necessarily representative for those of the global distribution of Galactic cosmic rays. Instead,  they could be determined by the peculiarities of recent injection of particles in the local ISM \citep{ptuskin06,horandel12,sveshnikova13,cream1,neronov12,pohl13,pohl13a,kachelriess15,kachelriess15_1}.

An illustration of these uncertainties is found in the  model of modification of the cosmic ray spectrum by the propagation effects encoded in an energy-dependent diffusion coefficient $D(E)\sim E^{\delta}$ \citep{Berezinsky_book}. The slope of the interstellar cosmic ray spectrum is determined by $\delta$ and by the slope $p_s$ of the injection spectrum, $p=p_{s}+\delta$. The most commonly considered acceleration mechanism is diffusive shock acceleration (DSA) \citep{Krymskii77,Bell78,Drury01}, which is expected to give a slope $p_s\simeq 2.0 ... 2.2$. Comparison with the slope of the locally measured cosmic ray spectrum, $p\simeq 2.7$,  points to a value of $\delta\simeq 0.5 ... 0.7$. This is, however, in tension with the measurements of the ratio of abundances of primary and secondary nuclei which give $\delta\simeq 1/3$ \citep{AMS-02_bc} and with the measurements of anisotropy of the cosmic ray flux \citep{blasi_anisotropy}. The value $\delta=1/3$ is also favoured by theoretical models of cosmic ray diffusion  through the ISM  \citep{Berezinsky_book}. 

Recent the data on cosmic ray positrons and antiprotons and also the  slope and anisotropy properties of the cosmic ray nuclei spectra suggest that the TeV range cosmic ray spectrum has a strong contribution a supernova which exploded approximately two million years ago within several hundred parsec distance from the Sun \citep{kachelriess15,kachelriess15_1}. The data on deposition of isotopes in the deep ocean crust  provide information on the occurrence of such nearby supernova events \citep{ellis96} indicate that only one such nearby supernova event has occurred over last 14 million years \citep{fields05,fry15}.  The slope of the locally observed cosmic ray spectrum in the TeV range might be largely determined by this recent supernova event. The information on the "characteristic" slope of the Galactic cosmic ray spectrum is completely erased by this last supernova contribution. 

Complementary information on the cosmic ray source and propagation parameters is provided by secondary \gr s and neutrinos.  Contrary to the charged cosmic rays,  neutral \gr s and neutrinos go straight from their production sites to the Earth. \gr\ and neutrino signal from individual sources could provide information on the injection spectrum of cosmic rays, while diffuse \gr\ and neutrino emission from the ISM could provide the data on the propagation of cosmic rays in the Galaxy. 

Analysis of the \gr\ signal from the Galactic Plane of the Milky Way Galaxy and from the disk of the Large Magellanic Cloud (LMC)  galaxy suggest that the average slope of the cosmic ray spectra produced by the star formation in these two galaxies is $p\simeq 2.45$, i.e. harder than that of the locally measured cosmic ray spectrum \citep{neronov15a,foreman15}. The slope of the neutral pion decay \gr\ spectrum produced by the interactions of cosmic rays with such hard slope is about $p_\gamma\simeq 2.4$. 

This slope is close to that of the slope of the spectrum of the isotropic diffuse \gr\ background (IGRB)\citep{fermi_EGB,fermi_EGB_new}.  This suggests that significant part of the IGRB might originate from the cosmic ray interactions in the star forming galaxies\citep{fermi_starforming,murase1,murase14} which are not individually resolved by the Fermi Large Area Telescope (LAT).  The normalisation of neutrino and \gr\ flux in the GeV energy range is fixed by the known relation between the far infrared and \gr\ luminosity of galaxies, $L_\gamma\simeq 10^{-4}L_{FIR}$ \cite{fermi_starforming}, and by the known level of extragalactic far infrared background, $F_{FIR}\sim 10^{-5}$~erg/(cm$^2$ s sr) \cite{EBL}. The most recent calculation of the \gr\ and neutrino flux from star forming galaxies which takes into account this relation along with the details of cosmological evolution of different types of  galaxies \cite{murase1,murase14} shows that they provide a significant contribution to the IGRB. 

An independent verification of the result on the hard $p\simeq 2.45$ average cosmic ray spectrum resulting from the star formation in the Milky Way and other galaxies could be obtained through the neutrino channel. The slope of the spectrum of neutrino emission from the Milky Way and other star forming galaxies is expected be about $p_\nu\simeq 2.4$, close to the slope of the \gr\ spectrum. 

The IceCube collaboration has recently reported the detection of astrophysical neutrino signal in the energy range from 10 TeV to 2 PeV \citep{IceCube_1yr,IceCube_PeV,IceCube_3yr,IceCube_2yr}. The signal forms a powerlaw spectrum $dN_\nu/dE \propto \left[E/100\mbox{ TeV}\right]^{-p_\nu}$,  with $p_\nu=2.46\pm 0.12$  \citep{IceCube_2yr}. 

This slope and normalisation of the neutrino spectrum are obviously inconsistent with the possibility that the observed neutrino flux is produced by cosmic rays in the Milky Way and / or other star forming galaxies, if one assumes that the typical spectrum of the cosmic rays resulting from the star formation is about the locally observed cosmic ray spectrum slope, $p\simeq 2.7$, and it has a knee feature exactly at the same energy as the knee of the locally measured cosmic ray spectrum \citep{Kachelriess2014}, an assumption was explicitly adopted in most of the previous calculations of the Galactic neutrino flux\cite{stecker79,berezinskii93,neronov0,neronov1,murase_ahlers}. Different assumptions about the average Galactic cosmic ray slope result in the estimates of the Galactic contribution to the astrophysical neutrino flux which differ by orders of magnitude (see e.g. \citep{ahlers15} for an example of calculation assuming $p=2.58$). However, none of the assumptions about the slope and position of the knee of the Galactic cosmic ray spectrum relying on local measurements is justified. 

The  measured slope of the astrophysical neutrino spectrum is consistent with the neutral pion decay \gr\ spectrum of the Milky Way and LMC disks \citep{neronov15a,foreman15}, as expected if the typical slope of the cosmic ray spectrum resulting from the star formation is $p\lesssim 2.5$ rather than $p\simeq 2.7$. Below we show that not only the slope, but also the normalisation of the neutrino flux is consistent with this hypothesis. 
We argue that this suggests the validity of a simple model in which the injection of cosmic rays with the spectrum with the slope suggested by the DSA, $p_{inj}\simeq 2 ... 2.2$  is followed by the softening by $\delta=1/3 ... 1/2$ produced by the escape through the turbulent galactic magnetic fields with Kolmogorov or Iroshnikov-Kraichnan turbulence spectrum. This simple model is valid for the spectrum of cosmic rays averaged over sufficiently large  regions of galaxies. It is, however, not valid for the single point measurements of the cosmic ray spectrum, such as the measurements at the position of the Solar system in the Milky Way.

\section{\gr\ and neutrino data analysis}

Our analysis uses all publicly available Fermi / Large Area Telescope (LAT) data \citep{atwood09} collected over the period from August 2008 till June 2014. We have processed the data using the  \textit{Fermi Science Tools v9r32p5}\footnote{http://fermi.gsfc.nasa.gov/ssc/data/analysis/} - a standard software package, provided by the Fermi collaboration to reduce the data, obtained by the Fermi/LAT. We have used the Pass 7 "reprocessed" event selection.

We have filtered the event lists using the {\it gtselect} tool with parameter  {\tt evclass=3}, which leaves the \gr\ events and rejects most of the cosmic ray background events. To produce the all-sky spectrum shown in Fig. 2, we have used the aperture photometry method\footnote{http://fermi.gsfc.nasa.gov/ssc/data/analysis/scitools/\\ aperture\_photometry.html}, applied to the full sky. An estimate of the exposure in each energy bin was done using the {\it gtexposure} tool with the option {\tt apcorr=no} (there is no need to correct the exposure for the point spread function for the full sky). To separate the diffuse emission from the point source contributions we subtract the flux in the circles of the radius $0.5^\circ$ around point sources from the four-year Fermi catalogue \cite{Fermi_catalog}. This is sufficient for the analysis of the diffuse emission in the energy band above 10~GeV. In this energy band the point spread function of the LAT is smaller than $\sim 0.3^\circ$ \cite{atwood09}.

We have verified that the aperture photometry approach gives the result which is consistent with the results obtained using the likelihood analysis. In particular, the spectrum of the $|b|>20^\circ$ part of the sky, calculated using the aperture photometry method is identical to that reported by  \cite{fermi_EGB_new} over the entire energy range from 100~MeV up to 1~TeV. 

To produce the Galactic latitude profiles for the neutrino data, we have  binned the neutrino events reported by \cite{IceCube_3yr} in Galactic latitude and have estimated the IceCube exposure in each Galactic latitude bin. This was done via a Monte-Carlo simulation of the Galactic latitude distribution of neutrino events driven from an isotropic sky distribution, with account of the declination dependence of the IceCube effective area, derived from the information reported by \cite{IceCube_3yr}. Having estimated the exposure in each Galactic latitude bin, we have multiplied the model fluxes of the Galactic and isotropic components by the Galactic latitude dependent exposure to estimate the expected number of neutrino counts in each latitude bin. 

\section{\gr\ and neutrino all-sky spectrum}

\begin{figure}
\includegraphics[width=\linewidth]{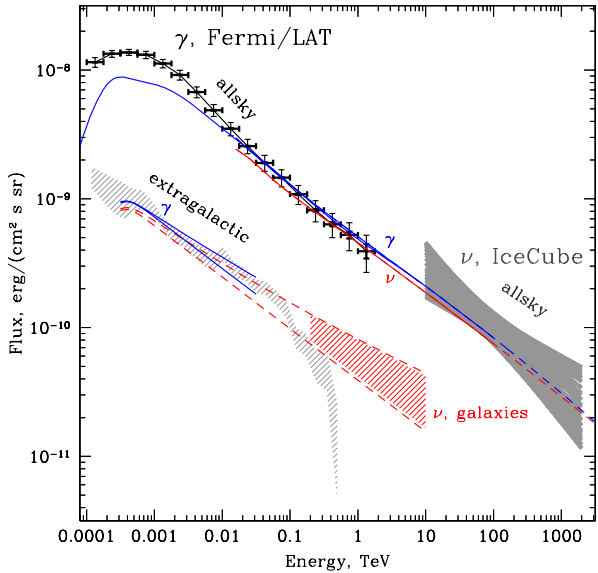}
\caption{Gamma-ray and neutrino spectra of the full sky.  Thick errorbars in the Fermi spectrum are for statistical error, thin errorbars are the systematic error. Red and blue thick solid curves show the neutrino and \gr\ emission from  protons with the  powerlaw spectrum with the slope $p=2.45$. Thin solid blue and dashed red curves show the estimates of possible \gr\ and neutrino fluxes from star forming galaxies.}
\label{fig:spectrum} 
\end{figure}

Fig. \ref{fig:spectrum} shows the combined \gr\ and neutrino all-sky spectrum in a broad GeV-PeV energy range. The statistical errors of the \gr\ signal are small in all energy bins up to $\sim 300$~GeV. The uncertainty of the \gr\ flux measurement is dominated by the systematic errors \citep{fermi_systematic}. The IceCube neutrino spectrum is derived from the analysis of \citep{IceCube_2yr}. The dark grey shaded band shows the 68\% uncertainty of the flux and slope of the neutrino signal.

From Fig. \ref{fig:spectrum} one could see that the neutrino spectrum lies at the high-energy extrapolation of the \gr\ spectrum of the entire sky. Thus, not only the slopes, but also the normalisation of the two spectra agree with each other. The uncertainty of the neutrino flux at 100~TeV is just by a factor of $\simeq 2$. Assuming a negligible uncertainty of the \gr\ flux at $\simeq 30$~GeV, one could find that a powerlaw fit to the combined \gr\ plus neutrino spectrum gives a very precise measurement of the slope of the powerlaw, $p_{\nu\gamma}=2.37\pm 0.05$, with an error $\Delta p_{\gamma\nu}=\mbox{log}(2)/\left[2\mbox{log}\left(100 \mbox{ TeV}/100\mbox{ GeV}\right)\right]\simeq 0.05$. This is due to a very large dynamic range of the energy on which the powerlaw is observed and to the moderate uncertainty of the neutrino flux measurement.

A more precise characterisation of consistency of the \gr\ and neutrino spectra is given by the model calculation shown in Fig. \ref{fig:spectrum}.  Red and blue thin solid curves show a model of the neutrino and \gr\ emission from interactions of cosmic rays with the ISM. The cosmic rays have a powerlaw spectrum with the slope $p=2.45$. The \gr\ and neutrino spectra are  calculated using the parametrizations of the pion production spectra by \cite{kelner,kamae06}.  The model does not fit the \gr\ data in the energy band below $\sim 10$~GeV. This is expected, because in this energy band significant contribution from electron Bremsstrahlung is expected  \citep{fermi_diffuse_2012,neronov15a}.

Consistency of the combined \gr\ and neutrino signal with a straightforward model of \gr\ and neutrino emission 
from a powerlaw distribution of the parent protons /nuclei suggests s simple model in which both the \gr\ and neutrino signal are generated by the interactions of cosmic rays produced by the star formation process. 

From Fig. \ref{fig:spectrum} one could immediately conclude that in such a scenario contribution of the Milky Way galaxy in the neutrino flux could not be negligible. Indeed, the neutrino flux from other star forming galaxies is constrained by the measurement of an upper limit on the \gr\ flux of star forming galaxies, which is given by the measured IGRB flux. Assuming that the \gr\ emission from cosmic ray interactions in star forming galaxies saturates the IGRB measurement, one could estimate the neutrino flux from star forming galaxies via extrapolation of the IGRB spectrum toward higher energies (the red hatched range in Fig. \ref{fig:spectrum}. This gives an upper limit of $\sim 50$\% of the contribution from star forming galaxies other than Milky Way. 

\section{\gr\ and neutrino all-sky anisotropy}
 
If a large part of the astrophysical neutrino signal is of Galactic origin, one expects to find higher signal level at low Galactic latitudes. The neutrino signal indeed shows a hint of anisotropy in the direction of the Galactic Plane \citep{IceCube_3yr}, which is consistent with the \gr\ -- neutrino signal correlation.    The distribution of neutrinos along the Galactic Plane is consistent with the distribution of the \gr s, with higher event statistics observed around the region of Galactic Ridge \citep{IceCube_3yr,neronov1}. 

To find the anisotropy properties of the Galactic component of the neutrino flux, we use the observed Galactic latitude profile of the \gr\ emission in the energy band above 300~GeV as a template. This is possible because Fig. \ref{fig:spectrum} suggests that the \gr\ and neutrino signals are both of hadronic origin and the neutrino signal above 10~TeV could be directly calculated from the \gr\ signal via s simple powerlaw extrapolation.  
In the energy band above 300~GeV the \gr\ signal is free from the extragalactic contribution which is suppressed by the effect of gamma-gamma pair production on the Extragalactic Background Light \citep{fermi_EGB_new}. Thus, the neutrino signal above 10~TeV is a sum of the anisotropic Galactic component which follows the \gr\ anisotropy template plus an isotropic extragalactic component. 

\begin{figure}
\includegraphics[width=\linewidth]{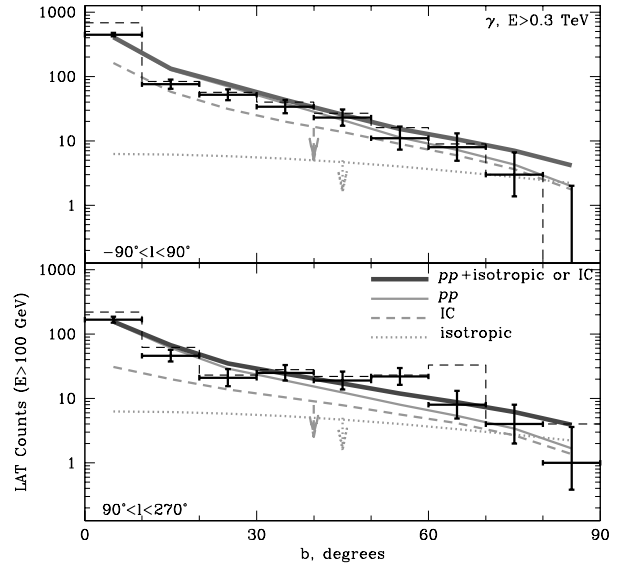}
\caption{Galactic latitude distribution of \gr s with energies above 0.3~TeV.  Upper and lower panels show the profiles for the inner and outer Galaxy. Dashed histogram shows the total counts in each bin, the data points show the profile after subtraction of counts from the sources in the 4-year Fermi catalog \cite{Fermi_catalog}.  The model curves for $pp$ and inverse Compton (IC) emission are taken from Ref. \cite{fermi_diffuse_2012}. }
\label{fig:profile_gamma} 
\end{figure}

 Fig. \ref{fig:profile_gamma} shows the Galactic latitude profile of the \gr\ signal compared to the "template" profiles of pion decay and inverse Compton (IC) emission calculated via powerlaw extrapolation of the model of \cite{fermi_diffuse_2012} to the energy band above 100~GeV. Assuming that the emission from the Galactic Plane (the first bin $|b|<3^\circ$) is dominated by the pion decay emission \citep{fermi_diffuse_2012,neronov15a}, one could find that the the pion decay component  provides a good fit for the profile in the whole latitude range up to $|b|=90^\circ$, without an additional contribution from the IC or isotropic flux. The sub-dominance of the IC contribution is also  supported by the modelling of the diffuse emission in the 100~GeV energy band in the Galactic latitude range $|b|>20^\circ$ reported by \cite{fermi_EGB_new}. This analysis finds that the IC component is sub-dominant even compared to the pion decay component calculated assuming the slope $2.7$ of the cosmic ray spectrum.
The dashed histogram in the two panels of Fig. \ref{fig:profile_gamma} shows the total flux in the latitude bins. One could  see that the point source contribution to the all-sky flux is minor.
 Thus, the observed Galactic latitude profile of the \gr\ flux above 300~GeV could serve as a proxy for the Galactic latitude profile of the Galactic component of the neutrino flux. 
 
\begin{figure}
\includegraphics[width=\linewidth]{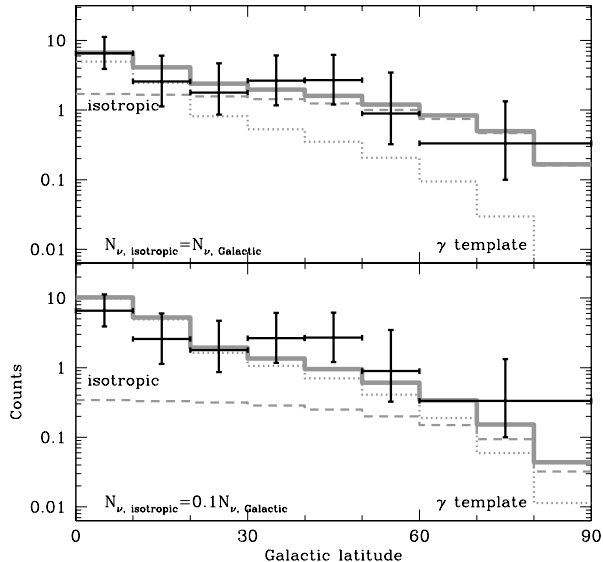}
\caption{Distribution of background-subtracted neutrino events with energies above 60~TeV in Galactic latitude (data points), compared with the distribution of $E>0.3$~TeV \gr s (thick grey histogram). Upper and lower panels show two models in which the isotropic contribution the neutrino flux (dotted histogram) is at the level of 50\% and 10\% of the Galactic neutrino flux, assumed to be proportional to the flux of $E>0.3$~TeV \gr s.}
\label{fig:profile_neutrino} 
\end{figure}

Fig. \ref{fig:profile_neutrino} shows the distribution of the neutrino events in Galactic latitude. It is compared to the model calculation taking into account the dependence of the IceCube exposure on the declination taken from the calculation of \cite{IceCube_3yr}. One could see that a range of models with large Galactic contribution to the neutrino flux (assumed to follow the Galactic latitude profile of the \gr\ flux above 0.3~TeV, shown in Fig. \ref{fig:profile_gamma}) is consistent with the data. The statistics of the signal is currently not sufficient for the discrimination of the Galactic and isotropic contributions.

 From Fig. \ref{fig:profile_neutrino}  one could see that the reduced $\chi^2$ of the fits of both Galactic latitude distribution models  the with 50\% and 90\% Galactic contribution is $\chi^2\lesssim 1$. The same would be true also for the isotropic models with 0\% Galactic contribution. Discrimination between the dominant Galactic or dominant extragalactic flux models will be possible already with IceCube in its present configuration. Fig. \ref{fig:profile_neutrino} shows that a decrease of the size of the errorbars by a factor of $\sim 2$ is needed for this. This could be achieved with the increase of signal statistics by a factor of 4, i.e. with a long exposure of IceCube, $\sim 12$~yr (compared to the available 3~yr).  Precision measurement of each component would, however, require a more powerful detector like IceCube-Gen2 \citep{gen2} or km3net \citep{km3net}. 

At the first sight, this result seems to be in tension with the analysis of \citet{ahlers15} who finds that the Galactic diffuse emission could not contribute more than 50\% of the observed neutrino signal. However, the discrepancy stems from model dependence of the results. One could notice that the result of \citet{ahlers15} depends on the choice of the spatial template for the hadronic component of the all-sky \gr\ flux. It relies on a spatial template derived by the Fermi collaboration \citep{fermi_diffuse_2012} assuming the reference cosmic ray spectrum with the slope 2.7 (rather than 2.58 assumed by \citet{ahlers15}). The uncertainty of this template and its dependence on the assumed spectral shape of the template are difficult to estimate.  In our analysis  we  assume that the Galactic longitude profile of the Galactic component of neutrino flux is identical to the profile of the \gr\ emission in the energy band above 300~GeV. This is motivated by the fact that the isotropic and the inverse Compton components of the \gr\ flux are suppressed at these energies. In such a way in our analysis we avoid using a model-dependent spatial template of the Galactic component of neutrino flux.

Detection of neutrino and \gr\ emission from the Galactic cosmic rays with $p=2.45$ powerlaw index provides an important clue for understanding of formation of Galactic cosmic ray spectrum \citep{kachelriess15,neronov15a}. It is consistent with the possibility that the cosmic rays are injected with the powerlaw spectrum with the slope $p=2.1 ... 2.2$ predicted by the shock acceleration models \cite{Krymskii77,Bell78,Drury01}, with the energy dependence of the diffusion coefficient $D(E)\sim E^{1/3}$ as expected for the Kolmogorov turbulence spectrum of the ISM \citep{armstrong95}.  

Although the slope and normalisation of the Galactic neutrino spectrum are fixed by the \gr\ data at lower energies, there is no certainty about its high-energy end. The Galactic cosmic ray spectrum should have a knee like suppression at high energies produced by the escape from the Galactic Disk and / or high-energy cut-off due to the absence of sources capable of particle acceleration beyond certain energy range \citep{Berezinsky_book,semikoz_knee,semikoz_knee1}. Similarly to the slope of the spectrum, the energy of the knee measured in the local Galaxy is not necessary the same as the average knee energy for the Galactic cosmic rays. The knee feature could be explained within the escape model of formation of cosmic ray spectrum as the  energy at which the scattering length of cosmic rays becomes comparable to the correlation length of the Galactic magnetic field \cite{ginzburg}. This model fits the shapes of the spectra of all elemental groups around the knee energy \citep{semikoz_knee,semikoz_knee1}. Within the escape model, the knee energy varies across the Galaxy and from one galaxy to another in response to the variations of the properties of the galactic magnetic field. Presence of the knee of the cosmic ray spectrum leads to a suppression of the neutrino flux at the energies $\sim 0.05$ of the knee energy. Uncertainty of the average position of the knee in the Galactic cosmic ray spectrum introduces an uncertainty in the shape of the spectra of Galactic \gr\ and neutrino emission in the PeV energy band (the dashed model line ranges in Fig. \ref{fig:spectrum}).

The locally observed TeV -- PeV cosmic ray proton spectrum appears "peculiar" in the sense that its slope is different from the slope of the average Galactic cosmic ray spectrum. This peculiarity could not be related to the specific of the propagation process in the local ISM, because in this case also the spectra of atomic nuclei would be affected. It could also hardly be related to the process of injection from the Galactic cosmic ray sources in general, because in this case the proton spectrum would be systematically softer than the nuclei spectrum everywhere in the  Galaxy and this would be visible in the softer \gr\ / neutrino spectrum of the Galaxy. 

The most probable reason for the peculiarity of the locally observed proton spectrum is in the change of the cosmic ray injection rate over the last $T\sim 10^7$~yr \citep{neronov12,pohl13,kachelriess15,kachelriess15_1}. Massive injection of cosmic rays in the local ISM in a star formation episode $10^7$~yr ago could have resulted in an increase of the cosmic ray energy  density at all energies. Faster diffusion of higher energy particles should have led to faster "wash out" of the excess density at high-energies and, as a consequence, to a temporary softening of the local  cosmic ray spectrum. Several candidate past events in the local Galaxy, like e.g. the event which produced an expanding ring of molecular clouds, the Gould Belt \citep{GB,neronov12a}, could be considered.  In a similar way, a nearby supernova explosion which occurred several million years ago could have produced an additional distortion in the cosmic ray spectrum. Evidence for such an event could be found through the analysis of the secondary positrons and antiprotons in the cosmic ray spectrum \citep{kachelriess15} and anisotropy \citep{kachelriess15_1}.  Discreteness of the distribution of the cosmic ray sources and time variability of the star formation rate generically leads to a situation in which the locally measured cosmic ray spectrum is not identical to the average Galactic cosmic ray spectrum  \citep{kachelriess15,kachelriess15_1,neronov15a}.

\section{Conclusions}

To summarise, we have shown that the the all-sky neutrino spectrum in the TeV-PeV energy range coincides with the high-energy extrapolation of the all-sky \gr\ spectrum in the energy band below 1~TeV. Such a consistency of the \gr\ and neutrino spectra is not unexpected in a simple model where both \gr s and neutrinos originate in the decays of neutral and charged pions produced by interactions of cosmic rays in the Milky Way and other galaxies. 

Noticing that the \gr\ all-sky spectrum in the energy band just below 1~TeV is dominated by the Galactic contribution, we were able to estimate the flux and anisotropy properties of the Galactic component of the neutrino spectrum. 

The isotropic neutrino flux from the cosmic ray interactions in star forming galaxies other than Milky Way is constrained by the level of the IGRB and could not constitute more than $\sim 50$\% of the observed neutrino flux. Presence of significant Galactic component of the neutrino flux (at least 50\% of the flux) is also consistent with the anisotropy properties of the neutrino signal. 

Combination of the \gr\ and neutrino data  suggests that the Galactic cosmic ray spectrum is harder than previously thought, with the slope $p\simeq 2.5$ in the TeV -- PeV energy range. The discrepancy between the slopes of the average Galactic cosmic ray spectrum and the locally measured cosmic ray spectrum is readily explained by the discreteness of the cosmic ray source distribution and time variability of the local star formation rate.

\end{document}